# Enhanced thermoelectric performance in TiNiSn-based half-Heuslers

R. A. Downie,[a] D.A. MacLaren,[b] R.I. Smith[c] and J.W.G. Bos[*a]

[a] *Institute of Chemical Sciences and Centre for Advanced Energy Storage and Recovery, School of Engineering and Physical Sciences, Heriot-Watt University, Edinburgh, UK, EH 14 4AS.*

[b] *School of Physics and Astronomy, University of Glasgow, Glasgow, UK, G12 8QQ.*

[c] *ISIS Facility, Rutherford Appleton Laboratory, Harwell Oxford, Didcot, UK, OX11 0QX.*

**Thermoelectric figures of merit, ZT > 0.5, have been obtained in arc-melted TiNiSn-based ingots. This promising conversion efficiency is due to a low lattice thermal conductivity, which is attributed to excess nickel in the half-Heusler structure.**

Thermoelectric waste heat recovery can be used to increase the efficiency of heat generating processes and is of great interest for a sustainable energy future. For example, thermoelectric power generators are considered for use in automobiles, where they recover heat from the exhaust gasses and replace the alternator, leading to an increase in fuel efficiency and reduced $CO_2$ emissions. Unfortunately, typical thermoelectric efficiencies are relatively low (~5%) and better performing materials are needed for this technology to find a wider application.

The thermoelectric efficiency of a material is given by a figure of merit; $ZT=(S^2/\rho\kappa)T$, where S is the Seebeck coefficient, $\rho$ the electrical resistivity, $\kappa = \kappa_{el} + \kappa_{lat}$ is the sum of the electronic and lattice thermal conductivities, and T is the absolute temperature.

Intermetallic half-Heusler thermoelectric phases have attracted considerable interest[1] but progress has been limited due to a large variation in reported properties. An example is the report of ZT = 1.5 for $Ti_{0.5}Zr_{0.25}Hf_{0.25}NiSn_{0.998}Sb_{0.002}$.[2] This was based on S = -350 µV K$^{-1}$, $\rho$ = 2 mΩ cm and $\kappa$ = 3 W m$^{-1}$ K$^{-1}$. Other groups have reproduced one or two of these favourable properties but never all three in the same material.[3-5]

The present study was motivated by the lack of structural information on these materials, which is an obvious starting point in probing the widely varying thermoelectric properties. The XYZ half-Heusler structure can be described as a face centred lattice of Z with X occupying the octahedral sites and Y on half of the tetrahedral sites.[6] This leads to a NaCl arrangement of X and Z and a Zincblende YZ sublattice. Band structure calculations support a Zintl-type electronic structure, where X transfers its valence electrons to the YZ sublattice.[7] For compositions with 18 valence electrons this leads to a closed shell electronic configuration and semiconducting behaviour.

We studied the nominally Sn deficient $Ti_{1-x}Zr_xNiSn_{0.95}$ (x = 0, 0.05, 0.50 and 1) series, as part of a wider investigation into the compositional flexibility of these materials. The structures were characterised using X-ray and neutron powder diffraction, and the thermoelectric figures of merit for $0 \leq x \leq 0.5$ were determined using physical property measurements. All samples were prepared by arc-melting. Immediately after arc-melting, the samples contained a large fraction of full-Heusler (e.g. $TiNi_2Sn$, where both tetrahedral sites in the face centred Sn lattice are occupied), Ti-Sn and Sn phases, which convert into the half-Heusler structure upon prolonged annealing at 900 °C inside vacuum sealed quartz tubes.[8]

An initial visual inspection of the X-ray diffraction patterns revealed sharp half-Heusler reflections, except for the x = 0.5 sample, which shows multiphase behaviour (Fig. S-1). We attribute the latter to the slow kinetics of mixing of Ti and Zr on the half-Heusler X-site. This is an additional poorly characterised feature of these materials which will be reported elsewhere.[9] Closer inspection of the diffraction data for the other compositions revealed subtle peak broadening, and two half-Heusler phases were used to obtain satisfactory profile fits. The refined lattice parameters for these phases are within $\Delta a = 0.06$ Å (Table S1), suggesting that these phases are closely related.

A complete characterisation of the thermoelectric figure of merit was undertaken (Fig. 1). The Seebeck coefficients show a similar temperature dependence with an asymptotic value of S = -175 µV K$^{-1}$ above 500 K. The temperature dependence of the resistivity is typical of semiconductors with a magnitude of 1 mΩ cm at high temperatures for all x. The maximum power factor ($S^2/\rho$) is a promising 3.5 mW m$^{-1}$ K$^{-2}$ at 650 K for x = 0 (Fig. S2). Similar magnitudes are observed for the other samples with the maximum shifting to higher temperatures for larger x. The thermal conductivity $\kappa \approx 4$ W m$^{-1}$ K$^{-1}$ for all samples. This is unexpected because the mass and size difference of Ti and Zr are effective impeders of the lattice thermal conductivity, and should therefore result in a decrease as x approaches 0.5. Stoichiometric TiNiSn is reported to have $\kappa = 7$-$8$ W m$^{-1}$ K$^{-1}$ at 300 K.[4, 8, 10] In particular, the thermal conductivities for x = 0 and x = 0.05 are therefore anomalously low.

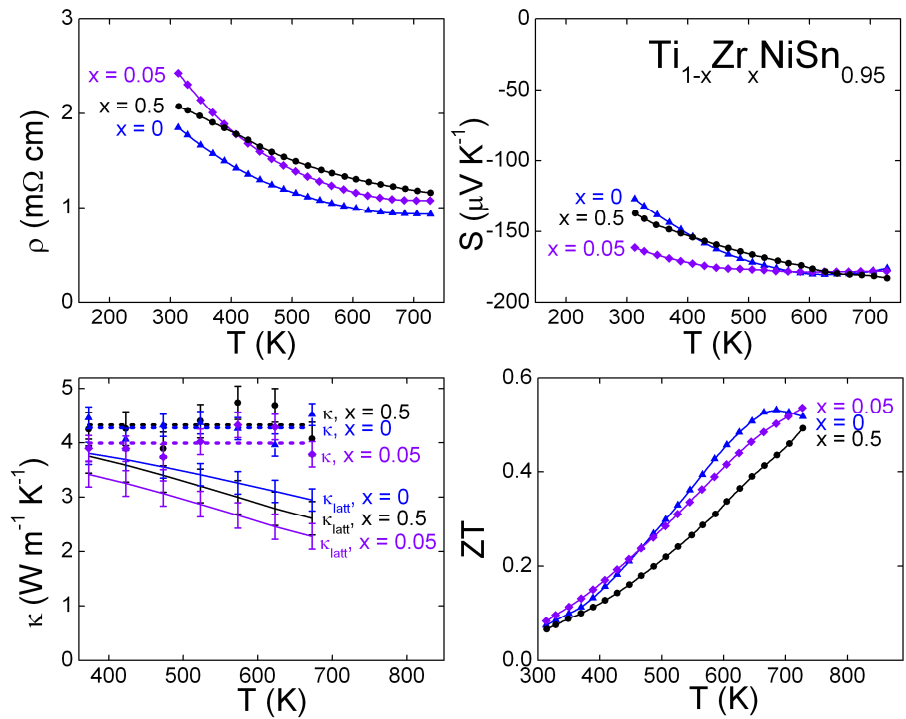

Fig. 1. Temperature dependence of the Seebeck coefficient (S), electrical resistivity (ρ), the total (κ) and lattice thermal conductivities ($\kappa_{lat} = \kappa - LT/\rho$; $L = 2.4 \times 10^{-8}$ W Ω K$^{-2}$), and the thermoelectric figure of merit (ZT) for the Ti$_{1-x}$Zr$_x$NiSn$_{0.95}$ series.

The neutron scattering contrast between Ti (-3.44 fm), Ni (10.3 fm) and Sn (6.23 fm) affords the experimental determination of the composition from refinement of the atomic site occupancies. Data were collected on powdered pieces of the x = 0 sample (~2 gram) using the newly-upgraded Polaris instrument at the ISIS facility, Rutherford Appleton Laboratory, UK. A multi-histogram Rietveld fit to Polaris detector banks 3-5 was done using the GSAS and EXPGUI suite of programmes.[11] The neutron data were fitted satisfactorily using a single phase, and the obtained results therefore represent a weighted average of the phases present.

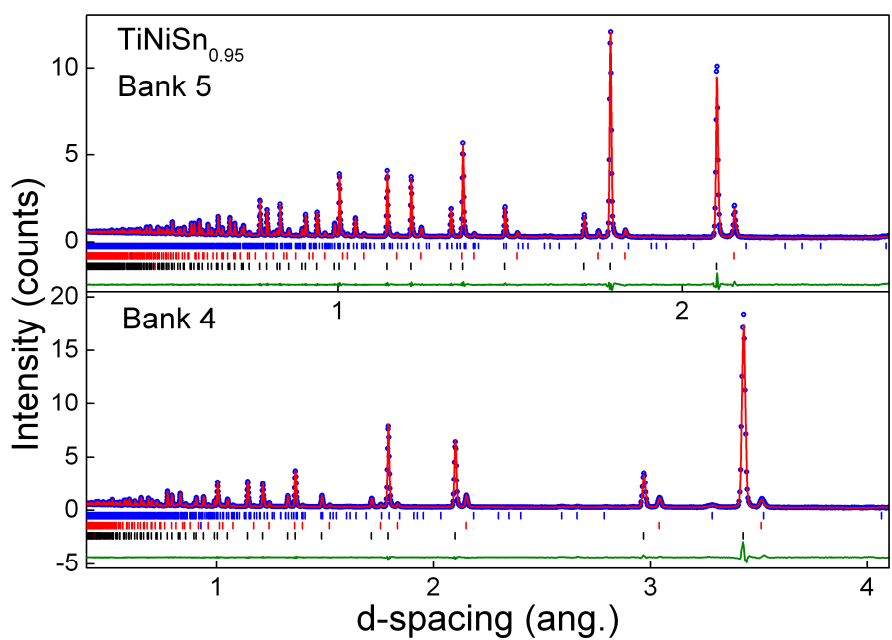

Fig. 2. Rietveld fit to Polaris neutron powder diffraction data for TiNiSn$_{0.95}$. The top markers (blue) are for Ti$_5$Sn$_3$; the middle markers (red) are for TiNi$_{1.78(1)}$Sn; the bottom markers (black) are for Ti$_{1.000(2)}$Ni$_{1.058(3)}$Sn$_{0.991(2)}$.

An initial refinement of the atomic site occupancies for TiNiSn$_{0.95}$ indicated an unrealistic composition of Ti$_{0.770(3)}$NiSn$_{0.842(2)}$ with a goodness of fit, $\chi^2 = 7.0$. Assuming stoichiometric TiNiSn ($\chi^2 = 11.2$) yielded large thermal displacement parameters on the Ti and Sn sites (3 and 2 times that of the Ni site, respectively), in keeping with the "observed" Ti and Sn site deficiencies. After some trial and error we observed that partial occupancy of the vacant Ni site in the half Heusler structure led to a more realistic composition: Ti$_{1.000(2)}$Ni$_{1.058(3)}$Sn$_{0.991(2)}$ with an identical goodness of fit ($\chi^2 = 7.0$). This indicates that the site occupancies in the half-Heusler structure are unity, and that there is a small amount of excess Ni on the vacant tetrahedral site. There is no evidence for substantial Sn deficiency or Ti-Sn inversion. This is in keeping with defect energy calculations that show that metal inversion on the NaCl sublattice is unfavourable.[12] The fitted structural parameters are summarized in Table 1 and the Rietveld fit is shown in Fig. 2. We also prepared and characterised a nominally stoichiometric TiNiSn sample, which yielded a composition of Ti$_{1.002(2)}$Ni$_{1.036(3)}$Sn$_{1.000(2)}$

(Table S2, Fig. S3), demonstrating that the excess Ni results from the synthesis method, and is not linked to the Sn deficiency in the starting composition, although this does afford a larger Ni excess.

Table 1. Refined structural parameters for the TiNiSn$_{0.95}$ sample from Rietveld fits against Polaris neutron powder diffraction data.

|     | Wyckoff | $x$  | $y$  | $z$  | occupancy | $U_{iso}$ (Å$^2$) |
|-----|---------|------|------|------|-----------|-------------------|
| Ti  | 4a      | 0    | 0    | 0    | 1.000(2)  | 0.00478(7)        |
| Ni1 | 4c      | 0.25 | 0.25 | 0.25 | 1.000(3)  | 0.00541(4)        |
| Ni2 | 4d      | 0.75 | 0.75 | 0.75 | 0.058(3)  | 0.00541(4)        |
| Sn  | 4b      | 0.5  | 0.5  | 0.5  | 0.991(2)  | 0.00434(5)        |

Space group F-43m, a = 5.9363(1) Å; Final fit statistics: $\chi^2$ = 7.0; Bank 5: wRp = 2.4%, Rp = 3.1%; Bank 4: wRp = 1.9%, Rp = 4.1%; Bank 3: wRp = 2.4%, Rp = 3.2%. The sample contains TiNi$_{1.78(1)}$Sn full-Heusler (9.7(1) wt%, a = 6.0803(1) Å) and Ti$_5$Sn$_3$ (6.7(3) wt%) impurities.

The diffraction results thus demonstrate that these samples contain excess Ni and are in some sense intermediate between the half-Heusler (TiNiSn) and full-Heusler phases (TiNi$_2$Sn). This is consistent with the two-phase behaviour observed in the X-ray diffraction data. Taking the x = 0 sample, lattice parameters of 5.929 Å and 5.935 Å were observed. The former is in very good agreement with the lattice parameter for TiNiSn prepared using conventional solid state reactions (5.930 Å).[9] The lattice parameter for TiNi$_2$Sn is 6.097 Å. Assuming that Vegard's law holds in the TiNi$_{1+y}$Sn "solid solution", this leads to y = 0.04(1) for the phase with the larger lattice parameter. The X-ray and neutron diffraction analysis are therefore in good agreement. A similar analysis can be made for the x = 1 sample, where the phases observed in X-ray diffraction correspond to ZrNiSn (a = 6.109 Å) and ZrNi$_{1.04(1)}$Sn (a = 6.115 Å).

The neutron data also reveal the presence of full Heusler and Ti$_5$Sn$_3$ phases with larger amounts observed for the Sn deficient sample. The Heusler phases are Ni deficient with identical refined compositions (TiNi$_{1.78(1)}$Sn). The presence of these phases is in keeping with their incomplete conversion into the half-Heusler structure. No elemental Sn was observed.

The impact of the excess Ni on $\kappa_{lat}$ is summarised in Fig. 3. The y = 0 data was taken from the literature (Ref. 4) as our stoichiometric sample prepared using conventional solid state reaction (Ref. 9) is only 75% dense, and no reliable $\kappa$ data can be obtained. The raw data for the y = 0.036 sample is shown in Fig. S4. A 40% reduction in $\kappa_{lat}$ is observed at 400 and 700 K. This is a dramatic decrease for such small doping levels (y ≤ 0.06). For comparison, in $Zr_{1-x}Hf_xNiSn$, x > 0.3 is needed to achieve a similar reduction.[3] The secondary $TiNi_2Sn$ and $Ti_5Sn_3$ phases are metallic with large $\kappa$, suggesting that the intrinsic $\kappa$ of the $TiNi_{1+y}Sn$ phase may be lower, and the impact of the excess Ni even more dramatic than observed here.

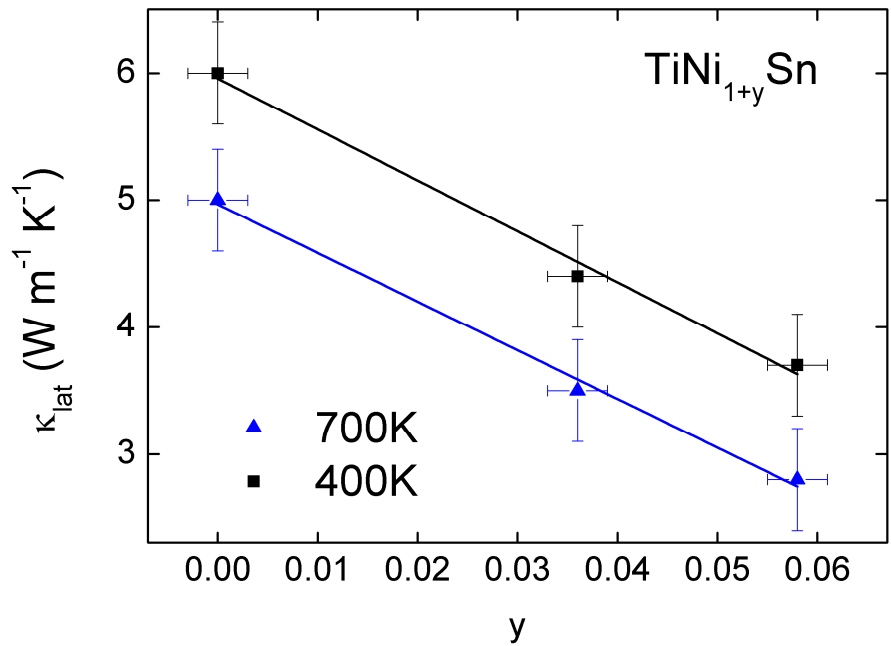

Fig. 3. The lattice thermal conductivity ($\kappa_{lat}$) for $TiNi_{1+y}Sn$. The y = 0 point was taken from Ref. 4. The lines are guides to the eye.

This study shows the link between synthesis protocol and physical properties and demonstrates that kinetic effects are important in the synthesis of these materials. The flexible Ni content of the Heusler structures is not widely appreciated. Comparing the present results and the literature suggests that prolonged annealing needs to be coupled to mechanical grinding to achieve a homogenous Ni dispersion. The present samples, where no homogenization of the ingots was performed after arc-melting, are therefore not at equilibrium. The advantage of the present route is that fully dense samples are obtained in an essentially single step method, and no further post synthesis densification is required. Note that even for homogenised samples minor full-Heusler and Ti$_6$Sn$_5$ impurities are observed.[13]

An unanswered question is how the Ni is dispersed within the sample. Neutron diffraction suggests that the excess Ni randomly occupies a small proportion of the vacant tetrahedra sites. However, diffraction probes an average and a solid solution between TiNiSn and TiNi$_2$Sn is not reported to exist.[14] It is therefore plausible that the excess Ni is clustered in small domains with a composition closer to the full-Heusler structure. The lattice parameter for TiNi$_2$Sn is much larger than for TiNiSn ($\Delta a$ = 0.15 Å). A full-Heusler inclusion would therefore introduce significant lattice strains. Under equilibrium conditions this is expected to result in complete phase separation but these samples form under kinetic control. A partial segregation is supported by recent electron microscopy work that reveals nano-sized Heusler inclusions and larger domains of TiNi$_{1.8}$Sn.[15] The presence of nano-inclusions is consistent with the large reduction in $\kappa_{lat}$ (Fig. 3) as the introduction of interfaces on this length scale is known to be an efficient source of phonon scattering.[16] The impact on S and $\rho$ is harder to estimate due to the polycrystalline nature of the samples and the presence of the TiNi$_2$Sn and Ti$_5$Sn$_3$ phases. One well understood effect in semiconductor nanocomposites with metallic inclusions is that of carrier filtering, which enhances S without increasing $\rho$ by filtering out low energy carriers.[17] The y = 0.04 sample (Fig. S4) has a lower S and a reduced $\rho$ compared to the y = 0 sample (S$_{300K}$ = -325 µV K$^{-1}$, $\rho_{300K}$ = 11 mΩ cm).[4] This suggests that the introduction of excess Ni does not result in carrier filtering but rather in conventional charge carrier doping. The y = 0.06 sample (x = 0 in Fig. 1) has further reduced S and $\rho$, consistent with conventional doping. However, an unambiguous interpretation of the electronic properties is complicated by the presence of varying amounts of metallic TiNi$_2$Sn and Ti$_5$Sn$_3$, which can also result in apparent decreases in S and $\rho$. The power factors (3.5-4.0 mW m$^{-1}$ K$^{-2}$) are enhanced compared to typical values for TiNiSn (2 mW m$^{-1}$ K$^{-2}$).[4] Combined with the low $\kappa$ this yields ZT values of 0.6 and 0.5 at 700 K for the y = 0.04 and y = 0.06 samples. These are very promising ZT values for "undoped" ternary TiNiSn.

To summarise, we have used neutron powder diffraction to demonstrate that arc-melted TiNiSn-based ingots naturally form with excess Ni in the half-Heusler structure. This results from kinetic constraints (spatial separation of reagents) in samples prepared without intermediate mechanical homogenisation. The excess Ni leads to low thermal conductivities and promising thermoelectric figures of merit (ZT > 0.5) at 700 K.

Supplementary Information for

**Enhanced thermoelectric performance in TiNiSn-based half-Heuslers**

R.A. Downie,[a] D.A. MacLaren,[b] R.I. Smith[c] and J.W.G. Bos[a]

Samples were prepared on a 3 gram scale from pieces of elemental Ti, Zr, Ni and Sn (all > 99.9% purity). Stoichiometric amounts were arc-melted three times under an Ar atmosphere. The ingot was turned over between melt attempts. No weight losses were observed. Following arc-melting, the ingots were annealed for two weeks at 900 °C inside vacuum sealed quartz tubes. The densities of the ingots were > 95%. X-ray powder diffraction patterns were collected over a period of 8 hours on a Bruker D8 Advance X-ray diffractometer with monochromoated Cu K$_{\alpha 1}$ radiation. The temperature dependence of the Seebeck coefficient and electrical resistivity was measured on a Linseis LSR-3 instrument. The thermal conductivity data were collected using an Anter Flashline thermal diffusivity instrument using a pyroceram reference sample.

Table S1. Refined lattice parameters (*a*), phase fraction (f), microstrain (S) and goodness-of-fit ($\chi^2$) for the arc-melted Ti$_{1-x}$Zr$_x$NiSn$_{0.95}$ phases.

| x | *a* (Å) | f | S (%)[a] | $\chi^2$ |
|---|---|---|---|---|
| 0 | 5.9293(1) | 0.34(1) | 0 | 2.9 |
|   | 5.9354(1) | 0.66(1) | 0.02(1) |   |
| 0.05 | 5.9459(2) | 0.16(1) | 0.01(1) | 2.6 |
|   | 5.9501(1) | 0.84(1) | 0 |   |
| 0.5 | 5.9909(1) | 0.23(1) | 0.70(1) | 2.4 |
|   | 6.0301(1) | 0.10(1) | 0.70(1) |   |
|   | 6.0612(1) | 0.67(1) | 0.30(2) |   |
| 1 | 6.1085(1) | 0.08(1) | 0 | 2.5 |
|   | 6.1148(1) | 0.92(1) | 0 |   |

[a] microstrain: S (%) = π/0.18 (LY-LY$_i$); LY$_i$ = 0.08°; GW fixed at 0.12°, LY$_i$ = instrumental contribution taken from x = 0 sample.

Fig. S1. Laboratory X-ray powder diffraction patterns for the arc-melted $Ti_{1-x}Zr_xNiSn_{0.95}$ samples after two weeks of annealing at 900 °C. Data were collected over a period of 8 hours on a Bruker D8 Advance diffractometer with monochromated Cu $K_{\alpha 1}$ radiation. Small amounts of full-Heusler (*) and $Ti_5Sn_3$ (#) remain after prolonged annealing.

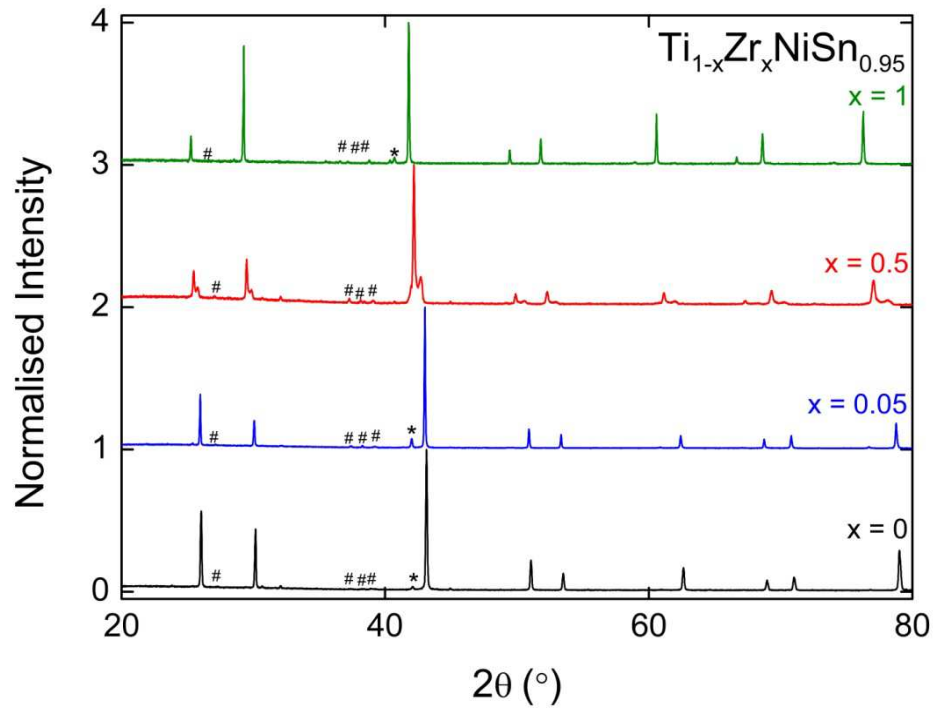

Fig. S2. Thermoelectric power factor ($S^2/\rho$) for the $Ti_{1-x}Zr_xNiSn_{0.95}$ samples.

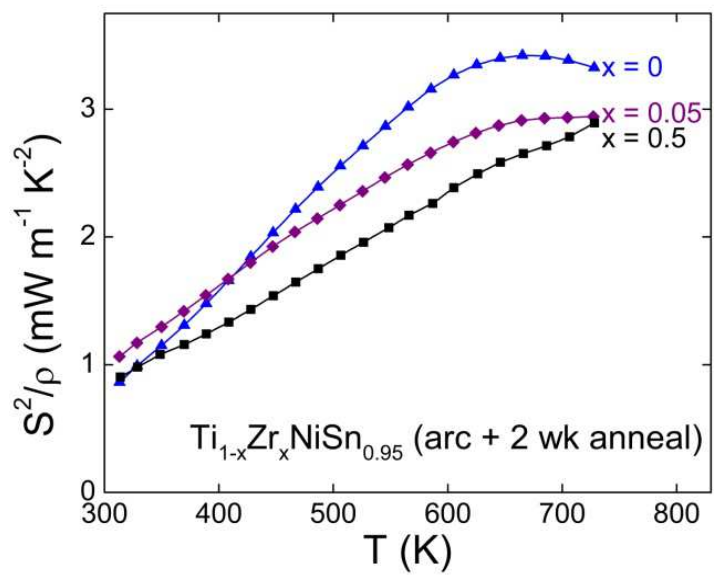

Fig. S3. Rietveld fit to Polaris neutron powder diffraction data for the nominally stoichiometric TiNiSn sample. Prepared by arc-melting and annealing for two weeks at 900 °C. Top markers (blue) are for $Ti_5Sn_3$ (2.3(1) wt%), middle markers (red) are for $TiNi_{1.76(2)}Sn$ (4.7(2) wt%), and bottom markers are for $TiNi_{1.036(1)}Sn$.

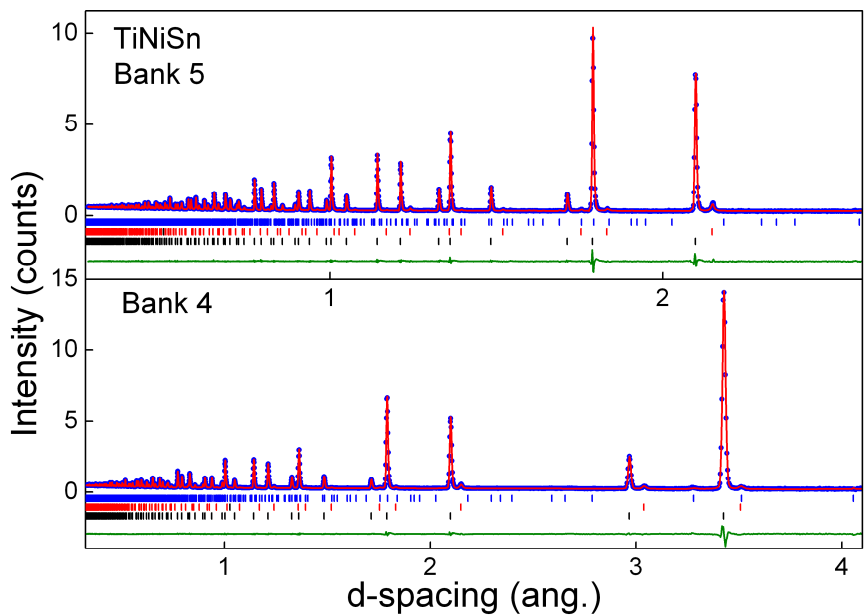

Table S2. Refined structural parameters for nominally stoichiometric TiNiSn from the Rietveld fit against Polaris neutron powder diffraction data.

|     | Wyckoff | $x$  | $y$  | $z$  | occupancy | $U_{iso}$ (Å$^2$) |
| --- | ------- | ---- | ---- | ---- | --------- | ----------------- |
| Ti  | 4a      | 0    | 0    | 0    | 1.002(2)  | 0.00478(7)        |
| Ni1 | 4c      | 0.25 | 0.25 | 0.25 | 1.000(3)  | 0.00525(5)        |
| Ni2 | 4d      | 0.75 | 0.75 | 0.75 | 0.036(3)  | 0.00426(5)        |
| Sn  | 4b      | 0.5  | 0.5  | 0.5  | 1.000(2)  | 0.00525(5)        |

Space group F-43m, a = 5.9327(1) Å; Final fit statistics: $\chi^2$ = 6.5; Bank 5: $wR_p$ = 2.5%, $R_p$ = 3.8%; Bank 4: $wR_p$ = 1.9%, $R_p$ = 3.6%; Bank 3: $wR_p$ = 2.5%, $R_p$ = 3.1%. The sample contains $TiNi_{1.76(2)}Sn$ impurity (4.7(2) wt%, a = 6.0753(3) Å) and $Ti_5Sn_3$ (2.3(1) wt%) impurities.

Fig. S4. Temperature dependence of the electrical resistivity (ρ), Seebeck coefficient (S), power factor ($S^2/\rho$), the total (κ) and lattice thermal conductivities ($\kappa_{lat} = \kappa - LT/\rho$; $L = 2.4 \times 10^{-8}$ W Ω $K^{-2}$), and the thermoelectric figure of merit (ZT) for the arc-melted nominally stoichiometric TiNiSn sample.

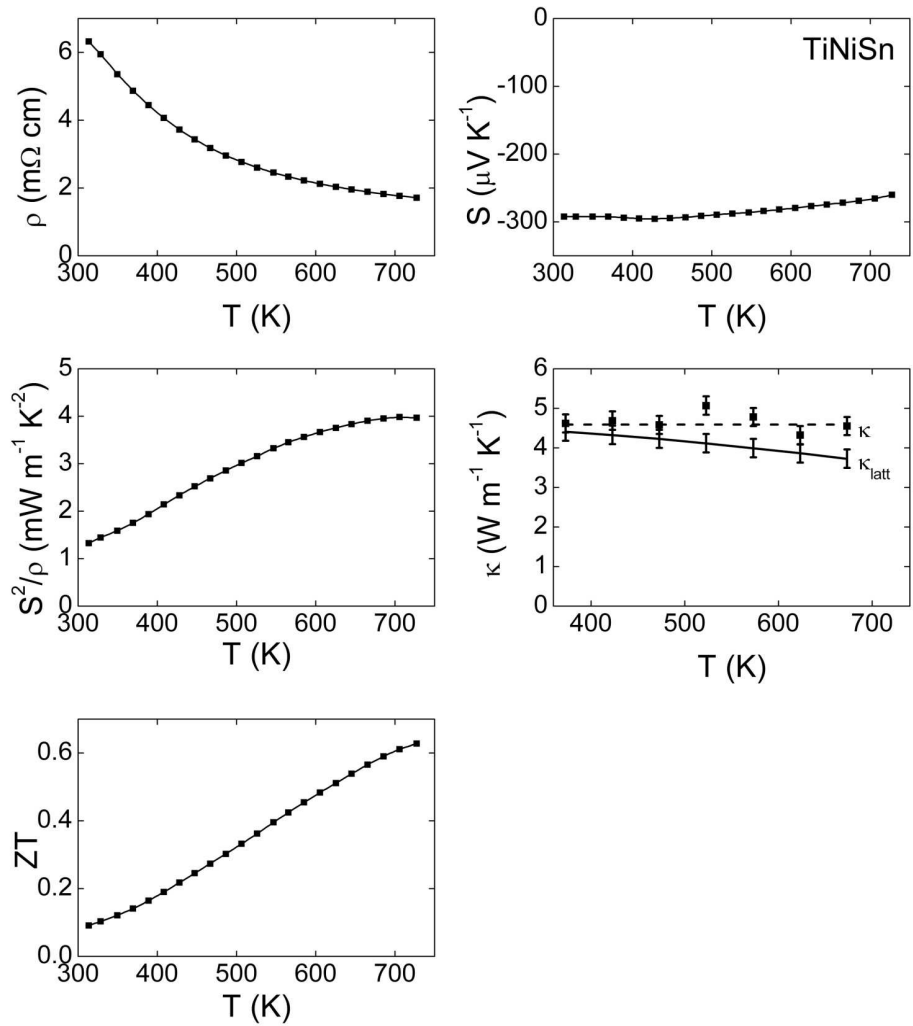